%% file: main.tex
\documentclass[conference]{IEEEtran}
\IEEEoverridecommandlockouts

\usepackage{balance}
\usepackage{cite}
\usepackage{amsmath,amssymb,amsfonts}
\usepackage{algorithmic}
\usepackage{graphicx}
\usepackage{textcomp}
\usepackage{xcolor}
\def\BibTeX{{\rm B\kern-.05em{\sc i\kern-.025em b}\kern-.08em
    T\kern-.1667em\lower.7ex\hbox{E}\kern-.125emX}}

\usepackage{pifont}
\usepackage{xcolor}

\usepackage[T1]{fontenc}
\usepackage{graphicx}
\usepackage{amsmath}
\usepackage{listings}
\usepackage{booktabs}
\usepackage{tcolorbox}
\usepackage{multirow}
\usepackage{hyperref}
\usepackage{xspace}
\usepackage{xcolor}
\usepackage{tikz}
\usepackage{amssymb}
\usepackage{xcolor}
\usetikzlibrary{positioning,arrows.meta}
\usepackage{cleveref}
\usepackage{soul}

\usepackage{etoolbox}

\usepackage{xcolor}     
\usepackage[normalem]{ulem} 
\usepackage{etoolbox}   
\usepackage{tcolorbox}  
\tcbuselibrary{breakable,skins} 

\definecolor{mygreen}{HTML}{117733}
\definecolor{myred}{HTML}{8A1414}

\lstdefinelanguage{dafny}{
  morekeywords={
    class, datatype,codatatype, type, iterator, lemma,
    bool, nat, int, object, set, multiset,seq, array,array2,array3, map,
    function, predicate,copredicate,
    pthost, var, static, refines,
    method, constructor,comethod,
    returns, yields, abstract, module, import, default, opened, as, in,
    requires, modifies, ensures, reads, decreases, free,
    match, case, false, true, null, old, fresh, choose, this,
    assert, assume, print, new, if, then, else, while, invariant, break, label,    return, yield, parallel, where, calc
  },
  sensitive=true,
  morecomment=[l]{//},
  morecomment=[s]{/*}{*/},
  morestring=[b]"
}

\lstdefinestyle{dafnystyle}{
 basicstyle       = \ttfamily\footnotesize,
  language        = dafny,
  keywordstyle    = \color{mygreen}\bfseries,
  commentstyle    = \color{gray}\itshape,
  stringstyle     = \color{red},
  frame           = single,
  numbers         = left,
  numberstyle     = \tiny\color{gray},
  breaklines      = true,
   columns         = fullflexible,
  showspaces      = false,
  showstringspaces= false,
  emph={Sorted,FindRange, Main,testRawsort,rawsort},
  emphstyle=\color{myred}
}

\lstdefinestyle{promptstyle}{
  basicstyle       = \ttfamily\scriptsize,
  language        = {},                
  breaklines      = true,
  breakatwhitespace = true,
  frame           = all,
}

\definecolor{grey}{RGB}{89, 89, 89}
\definecolor{blue}{RGB}{38, 105, 202} 
\definecolor{red}{RGB}{204, 102, 119} 
\definecolor{turquoise}{RGB}{68, 170, 153} 
\tcbset{
  myboxstyle/.style={
    colback=turquoise!5,
    colframe=turquoise,
    colbacktitle=turquoise,
    boxrule=0.4pt,
    arc=1mm,
    left=6pt,
    right=6pt,
    top=6pt,
    bottom=6pt,
    fonttitle=\bfseries,
    width=\columnwidth,
  }
}

\newcommand{\para}[1]{\par\smallskip\noindent\textbf{#1.}}

\newcommand{\parait}[1]{\par\smallskip\noindent\textit{#1.}}

\newcommand{\daisy}{\textsf{DAISY}\xspace}
\newcommand{\laurelfl}{DiagFL\xspace}
\newcommand{\laurelflB}{DiagFL+\xspace}
\newcommand{\llmfl}{LLMFL\xspace}
\newcommand{\llmexfl}{ExLLMFL\xspace}
\newcommand{\lgroundT}{ExactFL\xspace}
\newcommand{\Hybrid}{BothFL\xspace}

\newcommand{\haiku}{{\sc haiku}\xspace}
\newcommand{\opus}{{\sc opus}\xspace}

\newcommand{\enoex}{NoEx\xspace}
\newcommand{\erandom}{Random\xspace}
\newcommand{\etfidf}{TF--IDF\xspace}
\newcommand{\ecodeemb}{CodeEmb\xspace}
\newcommand{\emixemb}[1]{MixEmb$_{#1}$\xspace}

\newcommand{\nwoOne}{236}
\newcommand{\nwoTwo}{184}
\newcommand{\nwoAll}{28}
\newcommand{\nwComb}{448}

\AtBeginDocument{%
  \providecommand\BibTeX{{%
    Bib\TeX}}}

\title{DAISY: Fault-Guided Assertion Inference for Dafny \\
\thanks{\IEEEauthorrefmark{1} Author order follows a senior-author-last convention.}}

\author{
    \IEEEauthorblockN{Álvaro F. Silva}
    \IEEEauthorblockA{Faculdade de Engenharia\\
    Universidade do Porto \\ Porto, Portugal\\ alvaro.silva@fe.up.pt}
    \and
    \IEEEauthorblockN{Ruben Martins}
\IEEEauthorblockA{Computer Science Department\\
    Carnegie Mellon University \\Pittsburgh, USA\\
    rubenm@cs.cmu.edu}
    \and
\IEEEauthorblockN{Alexandra Mendes\IEEEauthorrefmark{1}}
    \IEEEauthorblockA{Faculdade de Engenharia\\
    Universidade do Porto \\ Porto, Portugal\\ alexandra@archimendes.com}
}
\begin{document}

\maketitle 

\begin{abstract}
Formal verification with Dafny provides strong correctness guarantees, but verification failures often require users to determine where helper assertions are needed and to manually write the assertions necessary for verification to succeed. This proof-engineering burden is difficult, time-consuming, and a barrier to the adoption of verification-aware languages.
We present \daisy (Dafny Assertion Inference SYstem), a tool for fault-guided assertion inference in Dafny. 
Given a Dafny program and a target method, \daisy localizes likely assertion sites using verifier diagnostics, LLM-based localization, or their combination. It then annotates the program with assertion placeholders, retrieves relevant in-context examples, generates candidate assertions with an LLM, and validates candidates by running the Dafny verifier.
To evaluate \daisy, we extend DafnyBench with verification-failing variants created by removing helper assertions from verified programs. The resulting benchmark distinguishes between programs with one missing assertion, two missing assertions, and all assertions removed, allowing us to evaluate \daisy on increasingly difficult proof failures. On this benchmark, \daisy verifies 86\% of programs with a single missing assertion and 36\% of programs with two missing assertions. These results show that fault-guided LLM assertion inference can repair many Dafny verification failures. They also show that successful repairs often differ from the original assertions, reducing proof-engineering effort without reconstructing the original proof.

\end{abstract}


\section{Introduction}

Verification-aware languages, such as Dafny \cite{leino2010dafny}, provide strong correctness guarantees by construction, enabling developers to annotate programs with specifications and automatically prove that the implementation satisfies them. In practice, however, verification rarely succeeds without additional effort: most proofs require numerous helper assertions\,---\,assertions that aid the verifier without affecting program behavior. Identifying which assertions are missing and where to place them is challenging. Reports from the industrial verification of cryptographic systems \cite{dodds2022} show that \emph{the majority of proof engineering effort is spent addressing failing proofs}, with missing assertions as a primary cause. Likewise, Oliveira et al. \cite{oliveira2025challenges} identify user challenges with writing annotations. 
This bottleneck makes assertion inference a crucial step toward improving the scalability and adoption of formal verification.

Large Language Models (LLMs) have recently been applied to reduce this burden by automatically inferring missing annotations \cite{mugnier2025laurel,poesia2024dafny}. Prior work shows that LLMs can generate assertions, loop invariants, program repairs, and full specifications \cite{mirchev2024assured,sun2024clover,wu2025specification}, lowering the manual effort of proof engineering. 
In particular, Laurel \cite{mugnier2025laurel} introduced a two-stage approach for Dafny: first localize a likely missing assertion location using verifier diagnostics, and then use retrieval-augmented prompting to generate an assertion. Laurel demonstrated that LLM-assisted assertion inference can unblock verification failures, but it primarily targeted lemmas and single missing assertions. This leaves open the broader problem of supporting verification failures in program code, where multiple helper assertions may be needed.

This paper proposes \daisy (Dafny Assertion Inference SYstem), a tool for helping developers resolve Dafny verification failures with helper assertions. \daisy preserves the original implementation and specification, searching only for auxiliary assertions that help the verifier prove the existing verification goals. By combining fault localization, retrieval-guided LLM generation, and verifier validation, \daisy identifies likely assertion locations and generates helper assertions to verify the Dafny program.
\daisy extends prior assertion-inference pipelines along three dimensions. First, it targets Dafny program code in addition to lemmas. Second, it supports inference for one or more assertion locations, enabling repairs that require multiple helper assertions rather than a single missing proof hint. Third, it provides a configurable retrieval component for selecting relevant in-context examples using different retrieval strategies. Together, these features make \daisy a practical framework for studying and deploying assertion inference for Dafny programs. 

To evaluate \daisy, we construct a new benchmark dataset of \nwComb{} programs by extending DafnyBench \cite{loughridge2024dafnybench} with verification-failing variants created by removing helper assertions from verified programs. The benchmark separates cases with one missing assertion, two missing assertions, and all assertions removed, enabling a controlled evaluation across increasingly difficult proof failures. \daisy verifies 86\% of single-assertion cases and 36\% of two-assertion cases. Many successful repairs differ from the original assertions, demonstrating that multiple valid proof strategies often exist.

Although our implementation targets Dafny, the underlying approach is not Dafny-specific. Many SMT-based verification tools rely on user-provided assertions, verifier diagnostics, and iterative proof debugging \cite{lattuada2023verus,filliatre2013why3,ahrendt2016deductivekeysmy,vazouliquidhaskell,astrauskas2019leveraging,swamy2016dependent}. The same combination of fault localization, example retrieval, LLM-based assertion generation, and verifier validation can therefore be adapted to other verification-aware languages.

\noindent
Our main contributions are:
\begin{itemize}
 \item \textbf{DAISY (Dafny Assertion Inference SYstem): a fault-guided assertion inference tool.}
  We present \daisy, a tool that repairs Dafny verification failures caused by missing helper assertions by inferring verifier-checked assertions while preserving the original implementation and specification. The tool and artifacts are available in \href{https://github.com/VeriFixer/daisy}{GitHub} with a self-contained artifact in \href{https://zenodo.org/records/20075857}{Zenodo}.

  \item \textbf{Fault localization for assertion insertion points.}
  We introduce and evaluate localization strategies that identify one or more likely assertion locations using verifier diagnostics, LLM-based prediction, or their combination.

  \item \textbf{Assertion generation for Dafny program code.}
    We extend assertion inference beyond lemma-centered settings to Dafny program code, and support configurable retrieval-guided generation with different strategies for selecting relevant in-context examples.

  \item \textbf{Benchmark and empirical evaluation.}
  We extend DafnyBench with 448 verification-failing variants containing one, two, or all assertions removed, enabling evaluation across proof failures of varying difficulty.

\end{itemize}

\section{Overview}
\input{figures/overview}

\daisy follows a two-stage pipeline inspired by Laurel~\cite{mugnier2025laurel}. 
First, \emph{fault localization} identifies likely locations where helper assertions are needed. 
Second, \emph{assertion inference} generates candidate assertions for those locations and validates them with Dafny. Throughout this process, \daisy preserves the original implementation and specification: the only modifications it attempts are the insertion of auxiliary assertions that help Dafny complete the proof. Figure~\ref{fig:daisy-overview} illustrates the pipeline, whose two main stages are described in detail in \S\ref{sec:method_fault_localization} (Fault Localization) and \S\ref{sec:assert_Inference} (Assertion Inference). 

We illustrate the pipeline with a concrete example. \lstinline[style=dafnystyle]!FindRange(q, key)!
takes a sorted sequence \lstinline[style=dafnystyle]!q! and returns the half-open interval \emph{[left, right)} containing exactly the elements equal to \lstinline[style=dafnystyle]!key!.
The example also uses the helper Boolean predicate \lstinline[style=dafnystyle]!Sorted(q)! that returns true if a given sequence is sorted in ascending order.

\begin{lstlisting}[
  style=dafnystyle,   
  numbers=none]
predicate Sorted(q: seq<int>){
  forall i,j :: 0 <=i<=j<|q| ==> q[i]<=q[j]}

method FindRange(q: seq<int>, key: int) 
 returns (left: nat, right: nat)
  requires Sorted(q)
  ensures left<=right<=|q|
  ensures forall i :: 0<=i<left ==> q[i]<key
  ensures forall i :: left<=i<right ==> q[i]==key
  ensures forall i :: right<=i<|q| ==> q[i]>key
\end{lstlisting}

In the method \lstinline[style=dafnystyle]!Main!, a call to \lstinline[style=dafnystyle]!FindRange! correctly computes the range of the value 10. However, Dafny's verifier cannot automatically deduce that the resulting indices are exactly 4 and 7. This failure to prove the assertion \lstinline[style=dafnystyle]!assert i == 4 && j == 7! indicates a need for additional helper assertions.

\begin{lstlisting}[
  style=dafnystyle,   
  numbers=none]
1: method Main(){
2:	var q := [1,2,2,5,10,10,10,23];
3:	assert Sorted(q);
4:	assert 10 in q;
5:	var i,j := FindRange(q, 10);
6:	assert i == 4 && j == 7 by {  /*Fails to prove*/
7:		assert q[0] <= q[1] <= q[2] <= q[3] < 10;
8:}}
\end{lstlisting}

\noindent\noindent

\daisy proceeds in two stages: localizing likely assertion locations and inferring assertions for them.

\para{Fault Localization}
\label{par:example_fault_loc}
\daisy first identifies candidate locations for helper assertions, using the combined fault localization strategy described in \S\ref{sec:method_fault_localization}. In this example, Dafny reports the diagnostic message \emph{``assertion might not hold''} at line~6, where the final assertion cannot be proved. The verifier-diagnostic strategy uses this message to select the position after line~5, while the LLM-based strategy selects the position after line~6. The combined strategy therefore considers both locations:

\begin{lstlisting}[
  style=dafnystyle,   
  numbers=none]
5      : var i,j := FindRange(q, 10);
after 5: /*<Assertion is Missing Here>*/
6      : assert i == 4 && j == 7 by {
after 6: /*<Assertion is Missing Here>*/
7      :   assert q[0] <= q[1] <= q[2] <= q[3] < 10;
\end{lstlisting}

\para{Assertion Inference}
\label{sec:example_motivation}
\daisy then prompts an LLM to generate candidate assertions for each placeholder, as described in \S\ref{sec:assert_Inference}. For each placeholder, the model returns a ranked list of up to 10 candidate assertions, which are verified in parallel. With sufficient CPU resources, validating ten candidates incurs approximately the same latency as validating one. Consequently, intermediate \texttt{pass@k} metrics have little practical significance, as users experience similar
latency for one or ten candidates. We therefore report only \texttt{pass@10}, which better reflects the tool's practical effectiveness. In this example, the first-ranked candidates for the two placeholders verify the program:

\begin{lstlisting}[
  style=dafnystyle,   
  numbers=none,
  emph={Sorted,FindRange, Main},
  emphstyle=\color{myred}]
5      : var i,j := FindRange(q, 10);
after 5: assert 10 in q[i..j]; /*Added*/
6      : assert i == 4 && j == 7 by {
after 6:   assert (j == |q| || q[j] > 10); /*Added*/
7      :   assert q[0] <= q[1] <= q[2] <= q[3] < 10;
\end{lstlisting}

Removing either added assertion causes verification to fail, showing that both assertions are needed. Notably, \daisy's solution differs from the original human-written assertions:

\begin{lstlisting}[
  style=dafnystyle,   
  numbers=none,
  emph={Sorted,FindRange, Main},
  emphstyle=\color{myred},firstnumber=5]
var i,j := FindRange(q, 10);
assert i == 4 && j == 7 by {
  assert q[0] <= q[1] <= q[2] <= q[3] < 10;
  assert q[4] == q[5] == q[6] == 10;/*Ground Truth*/
  assert 10 < q[7];/*Ground Truth*/} 
\end{lstlisting}

This example illustrates an important feature of \daisy: successful verification does not require reconstructing the original proof. A solution is valid whenever Dafny proves the original specification after the generated assertions are inserted. Thus, \daisy may produce different, and sometimes fewer, assertions than the human-written proof while preserving the same verification guarantees.

\section{Fault Localization}
\label{sec:method_fault_localization}

The first stage of our pipeline identifies where helper assertions should be inserted. Given a failing Dafny method and its verifier diagnostic, a localization strategy returns one or more candidate insertion points. We evaluate several strategies for this step: LLM-based localization, verifier-diagnostic localization, and a combined strategy that uses both. The resulting insertion points are converted into assertion placeholders and passed to the assertion inference stage described in \S\ref{sec:assert_Inference}.

\para{LLM-based Fault Localization}     
The LLM-based strategy prompts the model with the line-numbered method and verifier diagnostic, and asks it to return assertion insertion points. 
The prompt allows multiple positions, since a verification failure may require several helper assertions.
We evaluate two LLM-based configurations. The zero-shot configuration, \textbf{\llmfl}, uses only the target method and its verifier diagnostic, with no examples. The few-shot configuration, \textbf{\llmexfl}, augments the prompt with three dynamically retrieved examples that illustrate common assertion-placement patterns. These examples are selected using the retrieval strategy described in \S\ref{par:exampleSelection}, which is the best-performing strategy identified in \S\ref{sec:example_augmentation_anaysys}.

\label{heuristic-based-localization}
\para{Verifier-Diagnostic Fault Localization Strategy} This strategy localizes missing helper assertions by mapping Dafny verifier diagnostics to candidate insertion points. It builds on the diagnostic-based rules introduced in Laurel~\cite{mugnier2025laurel}, but extends them to cover additional verifier errors. Table~\ref{tab:heuristic} summarizes the diagnostic categories handled by each strategy and the insertion location associated with each category.

\daisy extends the Laurel-derived rules with four additional diagnostic categories. 
\emph{LoopInvariants} captures failures to prove that a loop invariant holds initially or is preserved by the loop body. \emph{TimeOut} denotes verification conditions that the SMT solver cannot discharge within the allotted time. \emph{SubsetConstraints} covers failures to prove that a value satisfies the constraints of its declared type. \emph{ElementNotInDomain} captures failures to establish membership in a required domain.

We refer to the strategy that uses only the Laurel-derived rules as \textbf{\laurelfl}, and to the extended strategy that uses all rules in the table as \textbf{\laurelflB}. 
Note that both strategies rely on the verifier's diagnostic messages and can return only \emph{one position} for generating an auxiliary assertion.

\label{hybrid-localization}
\para{Combined Fault Localization} 
We also evaluate a combined strategy, \textbf{\Hybrid}, that takes the union of the candidate positions returned by \textbf{\laurelflB{}} and \textbf{\llmexfl{}}. 
This method considers the union of assertion positions suggested by the heuristic-based strategy and those inferred by the LLM (an example was shown in the Fault Localization paragraph of 
\S\ref{par:example_fault_loc}). 
It combines the reliability of diagnostic heuristics with the contextual flexibility of LLMs, reducing erroneous placements relative to LLM-only localization, while offering greater positional diversity than purely heuristic-based methods.

\begin{table}[!t]
\centering
\caption{Error types and insertion strategies for \laurelfl{} and \laurelflB{}. \laurelfl{} uses only the rows marked with \emph{Laurel}. \laurelflB{} extends \laurelfl{} and also uses the rows marked \daisy. \\\textbf{B}: insert before the diagnostic line $l$ (i.e., at position $l-1$). \\\textbf{E}: insert at the end of the enclosing block.}
\begin{tabular}{llc}
\toprule
\textbf{Error type} & \textbf{Origin} & \textbf{Loc.} \\
\midrule
Assertion, Related, LHSValue, Calc, Constructed & Laurel~\cite{mugnier2025laurel} & B \\
Postcondition, AssertBy, Forall                  & Laurel~\cite{mugnier2025laurel} & E \\
LoopInvariants                                   & \daisy   & E \\
TimeOut, SubsetConstraints, ElementNotInDomain   & \daisy    & B \\
\bottomrule
\end{tabular}
\label{tab:heuristic}
\end{table}

In summary, for {\textbf{Fault Localization Strategies}} the following variants are explored:
\begin{itemize} 
\item \textbf{\laurelfl{}}: verifier-diagnostic localization using the Laurel-derived rules. 
\item \textbf{\laurelflB{}}: extended verifier-diagnostic localization using both the Laurel-derived rules and the additional rules introduced in \daisy. 
\item \textbf{\llmfl{}}: zero-shot LLM-based localization, using only the target method and verifier diagnostic. 
\item \textbf{\llmexfl{}}: few-shot LLM-based localization, adding three dynamically retrieved in-context examples. 
\item \textbf{\Hybrid}: combined localization using the union of positions returned by \laurelflB{} and \llmexfl{}. 
\end{itemize}

\section{Assertion Inference}
\label{sec:assert_Inference}

After fault localization identifies candidate assertion locations, \daisy passes an annotated version of the failing program to the assertion-inference stage. We call this annotated version the \emph{localized program}: the original failing Dafny program together with explicit placeholders at the insertion points predicted by the fault-localization stage. The role of assertion inference is therefore not to choose where assertions should be inserted, but to generate candidate helper assertions for the locations already selected by fault localization.

For each placeholder, the LLM generates a ranked list of candidate assertions.
During validation, \daisy constructs a bounded set of verification attempts from these ranked lists rather than enumerating the full Cartesian product of candidates across placeholders. We then evaluate each candidate individually by inserting it at its corresponding placeholder while leaving all other placeholders empty. We also evaluate aligned combinations by jointly inserting candidates with the same rank across all placeholders. This bounded search balances exploration of individual and combined repairs with a tractable number of verification attempts.

\para{Example Selection Strategies}
All assertion-inference configurations use the same localized program and the same assertion-generation prompt.
The only difference between the inference configurations is how the in-context examples are selected. In the zero-shot configuration, no examples are provided.
In all few-shot configurations, the prompt includes a few examples selected from a retrieval corpus of previous assertion-repair instances. In our experiments, we use a fixed retrieval budget of
\(k=3\) examples, although this is a configurable parameter of the tool. We exclude examples from the same functions, methods, or lemmas, to prevent data leakage. We organize the strategies into three groups: \emph{baselines}, \emph{lexical retrieval}, and \emph{embedding-based retrieval}.

\parait{Baselines} The first two strategies measure the effect of examples independently of retrieval quality.
\begin{itemize}
  \item \textbf{\enoex}: No in-context examples are included. This corresponds to a zero-shot prompt baseline.

  \item \textbf{\erandom}: Examples are selected uniformly at random from the retrieval corpus. 
\end{itemize}

\parait{Lexical Retrieval}
This strategy tests whether useful examples can be retrieved using explicit token overlap rather than learned semantic representations. This is a natural baseline for Dafny programs because similar proof failures often share concrete syntactic cues, such as assertion keywords, sequence operators, quantified expressions, helper lemma names, or recurring verifier-error phrases.
\begin{itemize}
  \item \textbf{\etfidf}: Few-shot prompt with the examples with the highest matching score compared with the target using a TF--IDF-based~\cite{ramos2003using} retrieval method.
  In our implementation, TF--IDF is computed over the Dafny method text itself. Following established practices in software engineering and code retrieval \cite{mugnier2025laurel}, we use TF–IDF as a classical baseline.

\end{itemize}

\parait{Embedding-based Retrieval}
\label{par:exampleSelection} 
The remaining strategies use semantic embeddings of the verifier diagnostic, the program text, or both. 
For each entry in the retrieval corpus, \daisy precomputes embeddings for the filtered verifier diagnostic and the method body using \texttt{jina-embeddings-v2-base-code}~\cite{gunther2023jina}. 
Although other embedding models could be used, we select this model because it is optimized for code-centric tasks and supports an 8,192-token context window, making it suitable for long method bodies and verifier traces.
For a new failing program, we embed its error and code, compute cosine similarities with all entries, and combine them as $\alpha \cdot \text{sim}_{\text{error}} + (1 - \alpha) \cdot \text{sim}_{\text{code}}$.
This approach is inspired by hybrid retrieval systems \cite{chen2025ownexploringoptimalembedding}, which integrate multiple semantic embeddings and may also incorporate keyword-based or other retrieval techniques.

\begin{itemize}
  \item \textbf{\ecodeemb}: Examples are selected using only code-embedding
  similarity. This corresponds to the case $\alpha=0$.

  \item \textbf{\emixemb{\alpha}}:  Examples are selected with $\alpha \in \{0.25, 0.5, 0.75, 1\}$. Larger values give more weight to
  the verifier diagnostic, while smaller values give more weight to code.
\end{itemize}

\section{Dataset}
\label{sec:dataset}

We built our dataset on top of DafnyBench~\cite{loughridge2024dafnybench}, a benchmark suite of 782 Dafny programs that successfully compile and verify. Starting from this verified corpus, we systematically remove helper assertions to create verification-failing variants, producing instances where our method must infer assertions sufficient to restore verification. Our goal is to evaluate assertion inference primarily in program code rather than only in proof-heavy settings. For this reason, DafnyBench is a natural starting point: unlike assertion-containing datasets such as DafnyGym~\cite{mugnier2025laurel}, which are dominated by lemmas and proof-oriented examples, DafnyBench contains a broader collection of Dafny functions, methods, and lemmas. We generated the following:

\begin{itemize}
  \item \textbf{w/o-1} (\emph{without one assertion}): We removed each assertion individually, one at a time. If removal caused verification to fail, we added both the missing assertion and the corresponding program to the \textbf{w/o-1} dataset. 
  
  \item \textbf{w/o-2} (\emph{without two assertions}): We removed two assertions simultaneously, considering all pairs containing assertions in the \textbf{w/o-1} dataset.  Pairs where one or both assertions were not part of \textbf{w/o-1} are excluded, as inferring them would be no harder than recovering a single assertion and therefore would not provide meaningful insight for our evaluation.

  \item \textbf{w/o-all} (\emph{without all assertions}): We removed all assertions, retaining only cases with at least three missing assertions to ensure no overlap with \textbf{w/o-1} or \textbf{w/o-2}. If verification fails, we add the missing assertions and the program to the \textbf{w/o-all} dataset.
\end{itemize}

After applying this procedure, we obtained \nwoOne{} \textbf{w/o-1}, \nwoTwo{} \textbf{w/o-2}, and \nwoAll{} \textbf{w/o-all} instances. The \textbf{w/o-all} dataset is small, with only \nwoAll{} cases, which is expected for several reasons. First, many assertions serve primarily as lightweight testing aids, so removing them eliminates the corresponding checks, causing the verification to pass simply because there is nothing left to verify. Second, the number of all-assertion removals is much smaller by
construction: a declaration with many assertions can produce many single- or
pair-removal instances, but only one all-removal instance. Finally, requiring at
least three removed assertions further restricts this category and keeps it
distinct from the \textbf{w/o-1} and \textbf{w/o-2} benchmarks.

Together, these benchmarks let us evaluate assertion inference across increasingly difficult proof failures: a single missing assertion, two interacting missing assertions, and cases where most local proof guidance has been removed.

\section{Evaluation}
We designed our evaluation around three research questions:
\begin{itemize}
\item \textbf{RQ1: How effective is \daisy for assertion inference?} We evaluate fault-localization and assertion-generation strategies on the \textbf{w/o-1}, \textbf{w/o-2}, and \textbf{w/o-all} benchmarks, showing that combining LLM-based and verifier-diagnostic localization performs best.

\item \textbf{RQ2: How effective is fault localization?} We compare our fault localization with the ground truth. Ground-truth positions reflect original developer assertions, but are not guaranteed to be optimal or minimal for verification.

\item \textbf{RQ3: Does example selection improve inference?}
We compare strategies for selecting in-context examples, including TF--IDF over program code, code-embedding retrieval, and multi-embedding retrieval over program code and verifier diagnostics.

\end{itemize}

\subsection{Experimental Setup}
\label{sec:experimental-setup}

We use the \nwComb{} Dafny programs (\nwoOne{} \textbf{w/o-1}, \nwoTwo{} \textbf{w/o-2}, and \nwoAll{} \textbf{w/o-all}) for assertion inference, generated as described in \S\ref{sec:dataset}. These programs fail to verify unless additional assertions are introduced. To ensure reproducibility, we fixed the toolchain to specific versions of Dafny \cite{leino2010dafny} (fork of v4.10) and Z3 \cite{mouraz3} (v4.15.4), with a 60-second verification timeout.

\para{LLM Selection}
To select the LLM used in our main experiments, we performed a pre-test on the assertion inference phase. We used a random subset of 30 benchmark instances and supplied each model with \textbf{\lgroundT{}} positions, i.e., the ground-truth location which corresponds to the original locations of the removed assertions in the verified programs. Evaluating all candidate models on the full dataset was computationally prohibitive, so we used this pre-test as a cost-benefit analysis. We compared several models (Claude Haiku/Sonnet/Opus 4.5, GPT 5.2, GPT 5 Mini, GPT 4.1, DeepSeek R1, Qwen3 Coder 480B/30B and Llama 3.3) through the Amazon Bedrock and OpenAI APIs in January 2026.

Considering both performance and cost, some models are dominated by others (lower accuracy at higher cost). After filtering, the most relevant candidates are Claude Opus 4.5, GPT-5.2, Claude Haiku 4.5, and Qwen3 Coder 480B.
Although \opus has the highest accuracy (73\%), its cost is substantially higher. For the core ablation studies, we focus on \haiku, which offers strong performance (67\%) at a lower cost (6.54 cents per run). \haiku slightly outperformed Sonnet on this small dataset, suggesting it is capable of detailed experiments, though this difference could be due to chance. The full ablation study with \haiku is estimated at around \$50, while running the same with \opus would scale to \$250, making it impractical. For the best methods, we also compare with \opus to assess differences with a higher-capacity model.

Each evaluated configuration combines a fault-localization strategy with an example-retrieval mechanism. Cost estimates are shown for Claude Haiku 4.5 (\$1 per 1M input tokens, \$5 per 1M output tokens), with default generation parameters.
Table~\ref{tab:costs} reports the estimated cost for evaluating one configuration on the full dataset. We compare our methods to a baseline, \textbf{FullResponse}, which queries the LLM for the complete output in a single step. This one-shot approach is not only less reliable, sometimes modifying unrelated code, but also more expensive, as larger outputs require more tokens. To allow a fairer comparison, \textbf{FullResponse(10)} requests 10 outputs per prompt, matching the number of alternatives generated by our methods. With the adjustments, FullResponse remains significantly more costly.

This two-stage separation, inspired by Laurel~\cite{mugnier2025laurel}, preserves verification soundness while enabling explicit evaluation of each stage, and lower costs than FullResponse prompting.

\begin{table}[!t]
\centering
\caption{Cost estimates for all evaluated configurations using Claude Haiku 4.5. Input/Output tokens are in millions (M) and thousands (K), respectively, considering Claude Haiku 4.5 base price}
\begin{tabular}{lrrrr}

\toprule
\textbf{Model} & \textbf{\# Prompts} & \textbf{Input (M)} & \textbf{Output (K)} & \textbf{Cost (\$)} \\
\midrule
\laurelflB{}        &       448 &        1.0 &        108 &        1.6 \\
\llmfl{}            &     896 &        1.6 &        109 &        2.1 \\
\llmexfl{}         &     896 &        2.2 &        109 &        2.8 \\
\Hybrid{}            &     896 &        2.2 &        109 &        2.8 \\
FullResponse (1) &       448 &        1.1 &        185 &        2.0 \\
FullResponse (10)&     4480 &       11.0 &      1850 &       20 \\
\bottomrule
\end{tabular}

\label{tab:costs}
\end{table}

\para{Fault Localization Evaluation}  
To evaluate the precision of the localization strategies, we categorize each predicted insertion point based on the verifier's response when the ground-truth assertion is inserted at that specific location. We define three mutually exclusive categories: 
\begin{itemize}
\item \textbf{Invalid}: where insertion causes a syntax error;
\item \textbf{Partial}: where insertion is syntactically valid but verification still fails;
\item \textbf{Valid}: where insertion is syntactically valid and verification succeeds when using the ground-truth assertion.
\end{itemize}

This classification moves beyond simple edit-distance metrics. In Dafny an assertion may be valid at several locations as long as the required facts are in scope and the assertion appears before it is needed. We observed in our experiments that many assertions have multiple valid locations, sometimes up to 25, and that valid locations for the same assertion can be separated by more than 40 lines.  

We compute the set of \textbf{Valid} positions using a perturbation procedure, where we move an assertion across all possible insertion points, and record the positions that restore verification. Because this procedure requires many verifier calls, we apply it only to \textbf{w/o-1} and \textbf{w/o-2}. For programs with two missing assertions, we fix one ground-truth assertion at its original location and move the other. We then repeat the process with the two assertions swapped. The union of all successful positions defines the valid-location set for that program. 

While this metric provides a robust evaluation framework, it is not exhaustive. The LLM may generate a different assertion, or a combination of assertions, that successfully verifies even if placed at a location where the original ground-truth assertion would fail. We analyze this in \S\ref{sec:rq2}.

\begin{table}[!t]
\centering
\caption{Verification success rate across benchmark categories using the best retrieval strategy, \emixemb{0.25}.}
\label{tbl:assertion-inference-verification}
\begin{tabular}{llcccc}
\toprule
\multirow{2}{*}{\textbf{Model}} 
& \multirow{2}{*}{\textbf{Localization}} 
& \multicolumn{4}{c}{\textbf{Benchmarks}} \\
\cmidrule(lr){3-6}
& 
& \textbf{w/o-1}
& \textbf{w/o-2}
& \textbf{w/o-all}
& \textbf{Combined} \\
\midrule

\multirow{6}{*}{\rotatebox[origin=c]{90}{\textsc{Haiku}}}
& \laurelfl & 52.5\% & 6.5\% & 10.7\% & 31.0\% \\ 
& \laurelflB & 57.6\% & 3.3\% & 10.7\% & 32.4\% \\ 
& \llmfl & 49.2\% & 9.8\% & \textbf{21.4\%} & 31.2\% \\ 
& \llmexfl & 53.4\% & 9.8\% & 17.9\% & 33.3\% \\ 
& \Hybrid & \textbf{67.4\%} & \textbf{14.1\%} & \textbf{21.4\%} 
& \textbf{42.6\%} \\ & \lgroundT & 71.6\% & 26.1\% & 25.0\% & 50.0\% \\
\midrule

\multirow{4}{*}{\rotatebox[origin=c]{90}{\textsc{Opus}}}
& \laurelflB   & 75.4\% & 6.0\% & 21.4\% & 43.5\% \\
& \llmexfl     & 78.0\% & 33.2\% & \textbf{28.6\%} & 56.5\% \\
& \Hybrid    & \textbf{86.0\%} & \textbf{35.9\%} & 21.4\% & \textbf{61.4\%} \\
& \lgroundT     & 91.1\% & 65.2\% & 50.0\% & 77.9\% \\

\bottomrule
\end{tabular}
\end{table}

\para{Statistical Testing} 
To assess whether performance differences between methods are statistically significant, we apply McNemar’s test with a significance level of $\alpha = 0.05$. We use this test to compare selected pairs of approaches, primarily contrasting LLM-based with heuristic-based methods, as well as comparing the best-performing method against alternatives.

\subsection{RQ1: How effective is \daisy for assertion inference?}
\label{sec:rq1answer}

To evaluate end-to-end assertion inference, we combine each fault-localization strategy with the best example-retrieval strategy, \textbf{\emixemb{0.25}} (identified in \S\ref{sec:example_augmentation_anaysys}). We measure whether the generated assertions restore verification on the \textbf{w/o-1}, \textbf{w/o-2}, and \textbf{w/o-all} benchmarks. As a reference point, we also include \textbf{\lgroundT{}}, 
which fixes the insertion points to the original assertion locations, isolating the effects of assertion generation and validation from localization.

We first evaluate all fault-localization strategies using \haiku. To control cost, we evaluate \opus only on the best methods: the best verifier-diagnostic strategy, \textbf{\laurelflB{}}, the best LLM-based strategy, \textbf{\llmexfl{}}, and their combination, \textbf{\Hybrid{}}. Table~\ref{tbl:assertion-inference-verification} summarizes the verification success rates. The ordering of results when using Both \haiku and \opus produce the same ranking. \textbf{\llmexfl{}} outperforms \textbf{\laurelflB{}}, and the best is \textbf{\Hybrid{}}.

Analyzing \haiku results in detail, we observe that \textbf{\laurelflB{}} improves upon \textbf{\laurelfl{}}, and that \textbf{\llmexfl{}} outperforms \textbf{\llmfl{}}. Overall, the performance of \textbf{\llmexfl{}} and \textbf{\laurelflB{}} is statistically indistinguishable. However, \textbf{\llmexfl{}} shows a notable improvement on the \textbf{w/o-2} dataset, increasing the success rate from 3\% to 10\%. 
The \textbf{\Hybrid{}} method consistently achieves the best overall performance among the evaluated localization approaches. Compared to \textbf{\laurelflB{}}, the improvement is statistically significant, as is the comparison with \textbf{\llmexfl{}}. 

With \opus, the ranking of methods remains consistent, with \textbf{\Hybrid{}} performing best across all benchmarks. Its improvements over \textbf{\laurelflB{}} and \textbf{\llmexfl{}} are statistically significant. Moreover, \textbf{\llmexfl{}} significantly outperforms \textbf{\laurelflB{}}, with the effect being particularly pronounced on the \textbf{w/o-2} dataset, where performance rises from 6\% to 33\%. 

\para{Multiple Missing Assertions}
Performance decreases as more assertions are removed. For example, OPUS \textbf{\Hybrid{}} verifies 86\% of \textbf{w/o-1} programs but only 36\% of \textbf{w/o-2} programs. This drop is larger than expected under an independence assumption: if the two assertions were inferred independently, the \textbf{w/o-2} success rate would be roughly  74\% $(0.86^2)$. The observed 36\% suggests that missing assertions are often interdependent, making multi-assertion inference substantially harder than solving two independent single-assertion tasks. Another contributing factor is our bounded pairing strategy. Rather than evaluating the full Cartesian product of candidate assertions, we consider only a limited set of pairings for computational efficiency. Consequently, a correct combination of assertions may exist but remain untested.

The \textbf{w/o-all} results are harder to interpret because this benchmark contains only 28 instances. In some cases, success does not decrease monotonically relative to \textbf{w/o-2}, likely because the small sample is sensitive to the particular difficulty of the selected programs. We therefore treat \textbf{w/o-all} as a stress test rather than as a primary basis for comparing methods. Still, the fact that several methods verify some \textbf{w/o-all} programs shows that restoring verification does not require reproducing the original proof structure exactly; alternative assertions and insertion points can also be sufficient.

\para{Effect of Combined Localization}
The strong performance of \textbf{\Hybrid{}} comes from combining complementary localization signals. Verifier-diagnostic localization is precise but limited to fixed error-message patterns, while LLM-based localization can exploit broader program context but may propose less reliable positions. Taking the union of both sets of positions creates a small portfolio of candidate locations that is more robust than either source alone.

\subsection{RQ2: How effective is fault localization?}
\label{sec:rq2}

\begin{figure}[!b]
\centering
\includegraphics[width=0.5\textwidth]{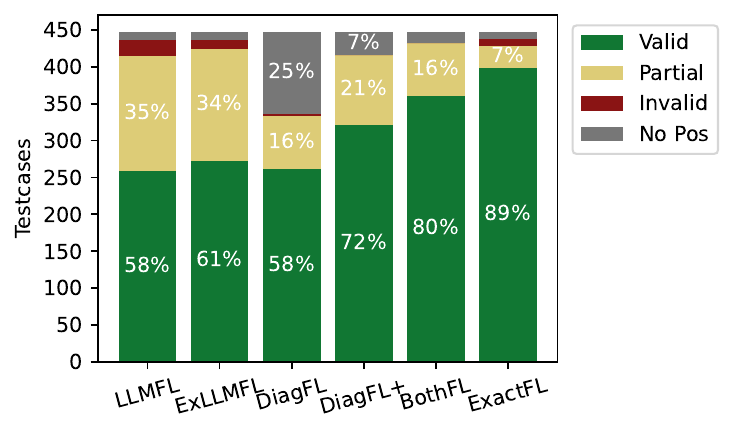}
\caption{Assertion insertion-position classification across localization strategies on the combined dataset of 448 Dafny programs using the \haiku model.}
\label{fig:fix-position-single}
\end{figure}

\para{Localization Quality}
We evaluate fault localization using the validity-based classification from \S\ref{sec:experimental-setup}.
Figure~\ref{fig:fix-position-single} summarizes the localization results. The  \textbf{\laurelfl{}} strategy leaves approximately 25\% of cases unsupported, mainly because it does not handle some verifier errors, such as loop-invariant failures. Our extended strategy, \textbf{\laurelflB{}}, reduces unsupported cases to 7\% and achieves the highest valid-position rate among individual methods, with 72\% valid placements. The LLM-based \textbf{\llmexfl{}} method also performs well, reaching 61\% valid placements. Overall, the combined \textbf{\Hybrid{}} strategy performs best, identifying valid placements in 80\% of cases.

\para{Combined Localization}
The strong performance of \textbf{\Hybrid{}} comes from combining complementary signals. Verifier-diagnostic localization is precise when the error message matches a supported pattern, but it can be too local or fail to return a position. LLM-based localization can use broader program context, but its predictions are less deterministic. Taking the union of both predictions gives a small portfolio of candidate locations that is more robust than either source alone.
The following example illustrates this complementarity:

\begin{lstlisting}[
  style=dafnystyle,
  numbers=none
]
method testRawsort() {
  var a : array<T> := new T[] [3, 5, 1];
  /*<Assertion is Missing Here>*/ ExLLMFL
  rawsort(a);
  /*<Assertion is Missing Here>*/ DiagFL+
  assert a[..] == [1, 3, 5];}
\end{lstlisting}

Here, \textbf{\laurelflB{}} places the missing assertion immediately before the failing check, following the verifier diagnostic. However, the useful assertion must be introduced before the call to \lstinline[style=dafnystyle]!rawsort(a)!, where it can relate the initial array contents to the subsequent sorting call. \textbf{\llmexfl{}} identifies this earlier location. By combining both predictions, \textbf{\Hybrid{}} avoids committing to a single localization strategy and reduces the risk of missing useful insertion points.

\para{Impact on Verification} Localization quality also affects end-to-end verification. We observed in our experiments that positions classified as \emph{valid} are 60\% more likely than \emph{partial} positions to lead to successful verification. This supports our validity-based metric as a more meaningful indicator of localization quality than distance to the ground truth.

\begin{table}[!t]
\centering
\caption{Verification success rate for example-selection strategies using \lgroundT{} fault localization and the \haiku model.}
\label{tbl:assertion-inference-verification-example}
\begin{tabular}{lcccc}
\toprule
\textbf{Example Selection}
& \textbf{w/o-1}
& \textbf{w/o-2}
& \textbf{w/o-all}
& \textbf{Combined} \\
\midrule
\enoex          & 59.3\% & 20.7\% & 14.3\% & 40.6\% \\
\erandom        & 59.7\% & 20.7\% & 17.9\% & 41.1\% \\
\etfidf         & 70.3\% & 23.9\% & \textbf{28.6\%} & 48.7\% \\
\ecodeemb       & 69.9\% & 23.9\% & 21.4\% & 48.0\% \\
\emixemb{0.25}  & 71.6\% & 26.1\% & 25.0\% & \textbf{50.0\%} \\
\emixemb{0.50}  & 68.2\% & \textbf{28.3\%} & 25.0\% & 49.1\% \\
\emixemb{0.75}  & \textbf{72.9\%} & 24.5\% & 21.4\% & 49.8\% \\
\emixemb{1.00}  & 68.2\% & 25.0\% & 17.9\% & 47.3\% \\

\bottomrule
\end{tabular}
\end{table}

\subsection{RQ3: Does example selection improve inference?}
\label{sec:example_augmentation_anaysys}

To isolate the impact of example extraction from fault localization, we evaluated several retrieval methods using the \textbf{\lgroundT{}} positions as a fixed baseline. In addition to standard baselines, we evaluate a hybrid retrieval strategy that combines code and error-message embeddings via a weighted cosine similarity metric, parameterized by $\alpha$ (\textbf{\lgroundT{}-\emixemb{\alpha}}).

Table~\ref{tbl:assertion-inference-verification-example} summarizes the results. \textbf{\enoex} and \textbf{\erandom} are statistically indistinguishable. This suggests that arbitrary examples provide little benefit. In contrast, all retrieval-based strategies improve over these baselines, showing that relevant examples are useful for assertion inference.
The best-performing method is \textbf{\emixemb{0.25}}, which achieves a 50\% success rate. This is a statistically significant improvement over \textbf{\enoex}, which only had a 41\% success rate. However, differences among retrieval-based methods are small. For example, \textbf{\emixemb{0.25}} does not significantly outperform the strongest non-\emixemb{} baseline, \textbf{\etfidf}. This suggests that retrieving relevant context matters more than the particular retrieval method.
We therefore used \textbf{\emixemb{0.25}} as the default example-selection strategy in the evaluation. It provides examples both for assertion inference and for the few-shot \textbf{\llmexfl{}} localization method.

\section{Discussion and Limitations}

Our best approach \textbf{\Hybrid{}} is unable to solve 14\% of the cases for the \textbf{w/o-1} dataset, and even if we had perfect fault localization, we would still have 9\% unsolved cases for this dataset, indicating that some assertions are inherently challenging to infer. We manually analyzed all \textbf{w/o-1} assertions that none of the methods could solve and identified categories that suggest directions for improvement.

\para{Failure Modes}
Several failures require proof structure beyond a single straightforward assertion. 
In some examples, the LLM correctly identified the required assertion but failed to provide a \texttt{by}-justification. For instance, in:
\lstinline[style=dafnystyle]! assert c <= x by { reveal A; } ! the model inferred \lstinline[style=dafnystyle]!assert c <= x;!, but did not recognize the need to reveal the named clause \lstinline[style=dafnystyle]!A! explicitly. A stronger integration of named assumptions and lemma references could likely solve such errors.

Many failures involved reasoning about sequence slicing or concatenation, which is common in Dafny proofs. Typical missing assertions include: \lstinline[style=dafnystyle]! assert rest == rest[0..1] + rest[1..];!, \lstinline[style=dafnystyle]! assert r == r[..k] + r[k..];!. Instead of producing these direct equalities, the model often attempted over-complicated multiset reasoning or type-incorrect variations. Improvements could come from incorporating explicit pattern templates for sequence equalities or augmenting the prompt with more examples of this type. 

\para{Context Limitations} \daisy prompts the model primarily with the target declaration, so relevant auxiliary functions, lemmas, types, or named assumptions may be omitted. Some assertions therefore depend on information unavailable to the model. Including surrounding definitions through targeted context retrieval could improve performance, but would increase prompt size, cost, and the risk of adding irrelevant context.

\para{Near Misses and Repair Opportunities}
Some failures were close to correct. For example, the model sometimes inferred the right algebraic structure but made a small arithmetic error, such as an off-by-one constant. 

\para{Evaluation Limitations}
As with other LLM-based evaluations, our results may be affected by model nondeterminism and potential data leakage from pretraining. Since each configuration was run once, the reported results may contain some sampling noise. However, repeated runs are often prohibitively expensive in this setting~\cite{mugnier2025laurel,loughridge2024dafnybench}. 
To reduce dependence on any single LLM output, each prompt requests up to 10 candidate assertions per placeholder. We therefore report verification success under a fixed 10-candidate budget: a benchmark instance is counted as solved if any verifier-validation attempt constructed from these candidates succeeds. 

Data leakage is another threat: models may have seen similar Dafny code or verification examples during pretraining, potentially inflating performance through memorization. We partially mitigate this concern with two observations. First, performance improves with prompt enrichment, such as adding retrieved examples, suggesting that model outputs depend on the provided context. Second, inferred assertions and insertion points often differ from the original ground truth, indicating that the model frequently synthesizes new verification strategies rather than simply reproducing memorized assertions.

\section{Related Work}

\para{Complete Program Synthesis}
Recent work integrates LLMs into Dafny verification workflows.  Clover~\cite{sun2024clover} generates code and specifications from natural language and checks their consistency with documentation and annotations. Other systems co-evolve programs, specifications, and tests~\cite{mirchev2024assured}, synthesize verified Dafny methods under different prompting strategies~\cite{misu2024towards}, or use verifier-guided search for multi-step Dafny programs~\cite{brandfonbrener2024}. Unlike these approaches, \daisy preserves the existing Dafny code and focuses on localizing and inferring helper assertions.

\para{Invariant Inference}
Invariant generation is a key subproblem in program synthesis for deductive verification, as invariants are essential for reasoning about loops \cite{floyd1993assigning}. Compared to full program synthesis, it is simpler and more tractable. Research in this area has achieved significant progress. For Dafny, Faria et al.~\cite{pascoal2025automatic} developed an LLM-based invariant inference tool that verifies 92\% of cases on the first attempt and 95\% within five attempts.

\para{Assertion Inference} 
The closest work to \daisy is Laurel~\cite{mugnier2025laurel}, which combines verifier-guided localization with LLM-based assertion generation for Dafny. Laurel identifies potential fix sites by analyzing Dafny's verifier error messages, applies heuristics to select promising locations, and retrieves similar examples to guide LLM prompting. Our work differs from Laurel
in several ways: (1) it evaluates a pipeline for multi-assertion inference and explores methods that directly generate multiple assertion candidates; (2) it uses LLMs not only for generation but also for predicting assertion insertion positions; (3) it shows that combining LLM-predicted locations with verifier-diagnostic localization yields better results; and (4) it introduces a framework for evaluating the quality of position inference independently of assertion generation.

\para{Proof Repair} Silva et al.~\cite{silva2024leveraging} studied GPT-4 for lemma discovery and proof sketching in Dafny, showing that even imperfect suggestions can help complete proofs. Poesia et al.~\cite{poesia2024dafny} inferred both loop invariants and helper assertions with a fine-tuned model, but rely on exhaustive testing across valid insertion points, which is costly for developer tools. Loughridge et al.~\cite{loughridge2024dafnybench} use LLMs to infill missing DafnyBench annotations, including loop invariants and assertions, reporting a 68\% success rate.

We also tried prompt-based infilling, but found it unreliable in our setting. Instead, \daisy first localizes likely assertion positions and then generates multiple candidate assertions in a single query. The difference in performance relative to DafnyBench-based infilling is partly due to dataset construction. Loughridge et al. evaluate on all removed annotations, whereas we keep only cases where removing an assertion causes a genuine verification failure. This excludes easier cases whose removal has no effect on verification, yielding a more challenging benchmark.

\section{Conclusion}
We presented \daisy, an LLM-assisted tool for inferring helper assertions in Dafny. \daisy separates assertion localization from assertion generation, enabling systematic evaluation of verifier-diagnostic localization, LLM-based localization, retrieval strategies, and model capabilities. We also introduced a curated DafnyBench-derived benchmark with variants where one, two, or all helper assertions are removed.

Our results show that combining LLM-based localization with verifier diagnostics yields the best overall performance. With combined localization and multi-embedding retrieval, \daisy verifies 86\% of programs with one missing assertion, 36\% with two missing, and 61\% overall. The results also show that successful verification does not require
reconstructing the original proof: inferred assertions and insertion points can differ from the human-written ones while still verifying.

Challenges remain for by-proofs, algebraic reasoning, multi-line assertions, and proofs requiring auxiliary definitions or lemmas. 
Still, DAISY
shows that fault-guided LLM assertion inference can repair a substantial fraction of Dafny verification failures, and the same localization-generation-validation structure may extend to other SMT-based verification frameworks.

\section*{Acknowledgments}
Álvaro Silva and Alexandra Mendes were partially funded by National Funds through the FCT - Fundação para a Ciência e a Tecnologia, I.P. (Portuguese Foundation for Science
and Technology) within the project VeriFixer, with reference 2023.15557.PEX (DOI: 10.54499/2023.15557.PEX). Álvaro Silva was also partially supported by Fundação para a Ciência e a Tecnologia (Portuguese Foundation for Science and Technology) through the Carnegie Mellon Portugal Program under the fellowship reference PRT/BD/155045/2024.

Ruben Martins was partially supported by the National Science
Foundation (NSF) under Award CCF2427581 and DARPA Agreement FA8750-24-9-1000.

Alexandra Mendes was also partially supported by an Amazon Research Award, Fall 2024.

\balance
\bibliographystyle{splncs04}
\bibliography{references}

\end{document}

%% file: figures/overview.tex
\begin{figure}[t]
\centering
\begin{tikzpicture}[
    >=Latex,
    font=\small,
    node distance=0.35cm and 0.65cm,
    stage/.style={
        draw,
        rounded corners=2pt,
        thick,
        align=center,
        text width=0.39\columnwidth,
        minimum height=1.15cm,
        inner sep=4pt
    },
    arrow/.style={->, thick}
]

\node[stage] (input) {
    \textbf{Dafny Program}\\[-1pt]
    {\scriptsize file + target method}
};

\node[stage, right=of input] (fl) {
    \textbf{Fault Localization}\\[-1pt]
    {\scriptsize likely assertion locations (\S\ref{sec:method_fault_localization})}
};

\node[stage, below=of fl] (gen) {
    \textbf{Assertion Inference}\\[-1pt]
    {\scriptsize candidate helper assertions (\S\ref{sec:assert_Inference})}
};

\node[stage, left=of gen] (verify) {
    \textbf{Verifier Validation}\\[-1pt]
    {\scriptsize Dafny-checked repair}
};

\node[stage, below=of verify] (output) {
    \textbf{\textcolor{mygreen}{\large{\checkmark}}\ Verified Program}\\[-1pt]
    {\scriptsize inferred assertions}
};

\node[stage, below=of gen] (fail) {
    \textbf{\textcolor{myred}{\large{\ding{55}}} Verification Fails}\\[-1pt]
    {\scriptsize try another candidate}
};

\draw[arrow] (input) -- (fl);
\draw[arrow] (fl) -- (gen);
\draw[arrow] (gen) -- (verify);
\draw[arrow] (verify) -- (output);
\draw[arrow] (verify) -- (fail);

\end{tikzpicture}
\caption{Overview of \daisy: localizing assertion locations, generating candidates, and validating them with Dafny.}
\label{fig:daisy-overview}
\end{figure}